# Surface Sensitive Microfluidic Optomechanical Sensors

Kyu Hyun Kim[1,2] and Xudong Fan[1,*]

[1]Department of Biomedical Engineering

University of Michigan

1101 Beal Ave., Ann Arbor, Michigan 48109, USA

[2]Department of Electrical Engineering and Computer Science

University of Michigan

1301 Beal Ave., Ann Arbor, Michigan 48109, USA

*Corresponding author: xsfan@umich.edu

**Abstract**

The microfluidic optomechanical resonator (μFOMR) based on a thin-walled glass capillary supports high Q-factor ($>10^3$) mechanical modes in the presence of liquids. In this Letter, the sensitivity of the μFOMR to the surface change is studied by layer-by-layer removal of $SiO_2$ molecules from the μFOMR inner surface using various concentrations of hydrofluoric acid solutions. A frequency downshift is observed with a sensitivity of 1.2 Hz/(pg/mm$^2$), which translates to a surface density detection limit of 83 pg/mm$^2$. This work opens a door to using the optomechanical mode for detection and characterization of molecules present near the resonator surface.

Discovery of optically excited mechanical vibrations in micro-optical resonators has broadened the scope of enhanced mechanical-matter interaction using light[1-4]. Light coupled into the whispering gallery mode (WGM) of a high Q-factor optical resonator transfers momentum to the wall of the resonator, exciting mechanical vibrations. Such excited mechanical modes are experimentally shown to be sensitive to mechanical properties and thus optomechanical resonators can be used as sensors for detection of inertia, pressure, mass, and temperature[5-9].

To enable optomechanical sensing in fluid, which is critical for biological, fluid dynamic, and rheological analyses, the sensing platform should be able to excite mechanical vibrations that can persist in and interact with liquids[10,11]. Unfortunately, most optomechanical sensors explored to date can only be implemented in air. When interacting with liquid, those sensors fail to oscillate due to the extremely high damping effect of liquids, which results in rapid loss of the mechanical energy from the sensor and hence low mechanical Q-factors. To overcome this issue, we recently developed a microfluidic optomechanical resonator (µFOMR) based on a thin-walled micron-sized glass capillary that is capable of exciting optomechanical vibrations with liquid flowing inside it[12,13] (Fig. 1a). Mechanical vibrations in the µFOMR with the frequency ranging from tens of MHz to tens of GHz and the mechanical Q-factor in excess of $>10^3$ have been achieved[12,13]. In addition, it is shown that the µFOMR could sensitively respond, in the form of the mechanical frequency shift and/or linewidth broadening, to the viscosity, fluid density, and pressure change inside the device[6,12-15]. The µFOMR becomes even more interesting considering the fact that the µFOMR shares nearly the identical thin-walled capillary structure that has been developed over the past 10 years under the name of the optofluidic ring resonator (OFRR) for sensitive label-free optical biosensing[16-20]. Combination of the µFOMR and the OFRR sensing mechanisms on the same platform would enable the dual-mode bioanalysis, which will provide complementary mechanical (mass, viscosity, and surface change, *etc*.) and optical (mass-induced refractive index and thickness change, *etc*.) information about biomolecular binding and conformational change on/near the capillary inner surface.

Towards this end, the first question would be the sensitivity of the μFOMR to the change on the capillary inner surface. In this Letter, we investigate the optomechanical frequency shift of the μFOMR in response to the sub-layer mass removal from the resonator surface. The results show that the μFOMR has a sensitivity of 1.2 Hz/(pg/mm$^2$) around an optomechanical frequency of 25 MHz, which translates to a minimally detectable mass density change of approximately 83 pg/mm$^2$.

The μFOMR used in this work is fabricated by stretching a fused silica capillary pre-form under $CO_2$ laser illumination, as described in detail previously[12,13,16,21]. During the fabrication, the $CO_2$ laser is modulated to produce a bottle-like capillary. The final μFOMR has an outer diameter of ~100 μm at its largest part and the wall thickness of ~10 μm (Fig. 1(a)). The experimental setup is shown in Fig. 1(b). The circular cross section of the capillary forms the optical ring resonator that supports the optical whispering gallery mode (WGM) traveling along the resonator circumference. Through an optical fiber taper the resonant light from a tunable diode laser (New Focus Velocity laser 6328, wavelength=1550 nm) is evanescently coupled into the optical WGM, which in turn excites the optomechanical modes of the capillary. It is reported that the threshold power for an optomechanical vibration of the μFOMR is on the order of 1 mW[12,13]. The optomechanical modes can be detected by a fast photodetector and analyzed by a spectral analyzer at the distal end of the fiber. The liquid is flowed through the capillary by a syringe pump. The optomechanical mode persists with a Q-factor in excess of 2000, even in the presence of liquid inside the capillary[12,13].

To quantitatively and systematically change the surface of the µFOMR, we employ the hydrofluoric acid (HF) etching method by flowing the pre-determined concentrations of HF solutions through the µFOMR. This method allows for layer-by-layer removal of $SiO_2$ molecules and has previously been used in characterization and precise control of microsphere and capillary based optical ring resonators[16,22,23]. Figure 2 plots real-time observation of the optomechanical frequency shift over time due to HF etching of the µFOMR inner surface using 1.25% (v/v) HF solution. A total of ~240 kHz frequency downshift is achieved in 23.5 minutes, suggesting that the µFOMR is capable of sensing the change on/near the resonator surface. During the etching process, the high density (2.2 $g/cm^3$) $SiO_2$ layer is replaced by the low density (1.0 $g/cm^3$) water (or 1.25% HF in water - to be exact) layer, resulting in a change in both the resonator geometry (such as the ratio of the outer diameter and inner diameter) and the effective mass experienced by the mechanical mode, and hence a shift in resonance frequency. Note that during the etching the mechanical Q-factor remains at approximately 2000.

To further investigate the effects of the etching on the resonance frequency, various concentrations of HF solutions are used. Since silica etching is an irreversible procedure, a new µFOMR needs to be used for each etching experiment. We therefore select a series of the µFOMRs with similar geometries and resonance frequencies (between 20-27 MHz). Fig. 3(a) shows the corresponding frequency shift. As expected, a faster shift of resonance frequency is observed for a higher HF concentration. For comparison, when water (0% of HF) is used, no frequency shift is observed over time. In order to compare the etching effect directly, fractional shift of resonance frequency, $\Delta f/f_0$, where $\Delta f$ and $f_0$ are the frequency shift and the resonance frequency, respectively, is calculated and plotted in Fig. 3(b). $\Delta f/f_0$ is related to the etching process by:

$$\frac{\Delta f}{f_0} = \frac{-\Delta E}{2E_{total}}, \quad (1)$$

$\Delta E$ and $E_{total}$ refer to as the energy change resulting from the mass removal from the μFOM inner surface and the total energy stored in the μFOM, respectively. The inset of Fig. 3(b) shows that the fractional shift follows the linear etching rate with respect to the increase in the HF concentration.

Institutively, removing mass from a mechanical resonator would lead to an increase in resonance frequency. However, our results indicate the opposite. In fact, the direction of the frequency shift depends not only on mass, but more importantly, on the mechanical energy distribution within the resonator, as revealed by Eq. (1) and Ref. 14. Figure 4 shows Finite Element Method (FEM) mode analysis of the μFOMRs. Here, resonance frequencies are calculated for three modes as the inner wall is thinned while water is inside the resonator. Gradual removal of inner layer induces resonance frequency downshift as experimentally observed. By comparing the frequency shift resulting from the thickness change to the experimental data, we determine that the breathing mode (2,1,0) is likely to be used in our experiment.

To estimate the sensitivity and detection limit of the mass change near the μFOMR inner surface, we use the etching curve for 1.25% HF solution. According to Fig. 3(a), in 23.5 minutes, the mechanical frequency is down-shifted by 240 kHz. Based on the etching rate of 7 nm/min[24], we arrive at a surface mass sensitivity of 1.2 Hz/(pg/mm$^2$), similar to the theoretical result reported on micro-toroid based optomechanical resonators in air[8]. Using the standard deviation of approximately 100 Hz achieved experimentally in our frequency detection[13], the surface density detection limit of about 83 pg/mm$^2$ can be realized. Furthermore, considering the surface area of $5 \times 10^{-2}$ mm$^2$ (80 µm inner diameter and ~200 µm in length based on the FEM result in Fig. 4), the mass detection limit is on the order of 4 pg, which is equivalent to a 2 µm diameter silica micro-particle deposited on the µFOM inner surface. The above mass detection results are similar to what has been experimentally achieved in polyethylene micro-particle detection using the fundamental optomechanical mode of a micro-toroid in air[8] and to the simulation results on the µFOM[14].

In summary, we have reported the µFOMR's capability in detecting mechanical property changes near its inner surface. The sensitivity of 1.2 Hz/(pg/mm$^2$) and the detection limit of 83 pg/mm$^2$ are achieved, similar to those obtained with micro-toroid based optomechanical resonators in air, which makes the µFOMR a highly sensitive mechanical sensor in liquid environment. In the near future, we will implement commonly used optical detection techniques such as laser wavelength stabilization and differential detection that can cancel out drifts in resonance frequency to increase the frequency detection resolution. Meanwhile, higher orders of harmonics (rather than the fundamental mode) and the modes with higher sensitivities (such as wineglass mode shown in Fig. 4) will be employed to increase the sensitivity. Together, we expect 10-100 fold improvement in the µFOMR's detection limit (*i.e.*, 1-10 pg/mm$^2$), which makes the surface sensing performance of the µFOMR comparable to that for the majority of optical label-free biosensors[19], in particular, the OFRR based ones[18]. Consequently, it becomes possible to carry out simultaneous mechanical and optical measurements using the same capillary device, in which the mechanical and optical modes work in concert to provide more detailed information of (bio)molecules near the surface.

The authors thank Qiushu Chen and Hengky Chandrahalim for µFOMR fabrication and discussion, and the National Science Foundation (CMMI-1265164) for support.

**Captions**

Figure 1  (a) Picture of a μFOMR. The device is made by stretching a fused silica pre-form under $CO_2$ laser illumination. The $CO_2$ laser intensity is modulated to generate curvature along the capillary. The μFOMRs used in this work has a typical outer diameter of ~100 μm with the wall thickness of ~10 μm. Inset: Cross-sectional view of the μFOMR. The circular cross section supports the optical whispering gallery mode that excites optomechanical vibrations. (b) Experimental setup.

Figure 2  Optomechanical resonance frequency shift over time due to the etching with 1.25 (v/v %) HF solution. At initial state, the resonance frequency is 20.68 MHz (peak #1). With the introduction of HF solution, the frequency down-shifts, as consecutively shown in peak #2 - #5 for 6 minutes, 12 minutes, 20 minutes, and 23.5 minutes of etching, respectively. The total shift is 240 kHz. During the measurement, the flow rate of HF solution is kept constantly at 0.08 mL/min.

Figure 3  (a) Absolute optomechanical resonance frequency shift over time extracted from Figure 2 for different concentrations of HF solution. Solid lines are the linear fit. (b) Fractional resonance frequency change derived from (a). Solid lines are the linear fit. Inset: Slope of fractional resonance frequency change as a function of HF concentration. Solid line is the linear fit.

Figure 4  Optomechanical modes and their resonance frequencies of the μFOMs are calculated as the wall thickness of the resonator is decreased from inside. Modes

calculated in this work are two breathing modes, (N,L,M)=(1,1,0) and (2,1,0), at resonance frequency of 17.94 MHz and 27.69 MHz, respectively, and a wineglass mode (1,1,5) at resonance frequency of 24.15 MHz. N: radial order; L: axial order; M: azimuthal order. Insets: The corresponding mode shapes. Parameters used in simulation: Inner diameter=79 µm, wall thickness=10.5 µm, curvature along the capillary=16000 µm. Outside: air; Wall: silica; Inside: water.

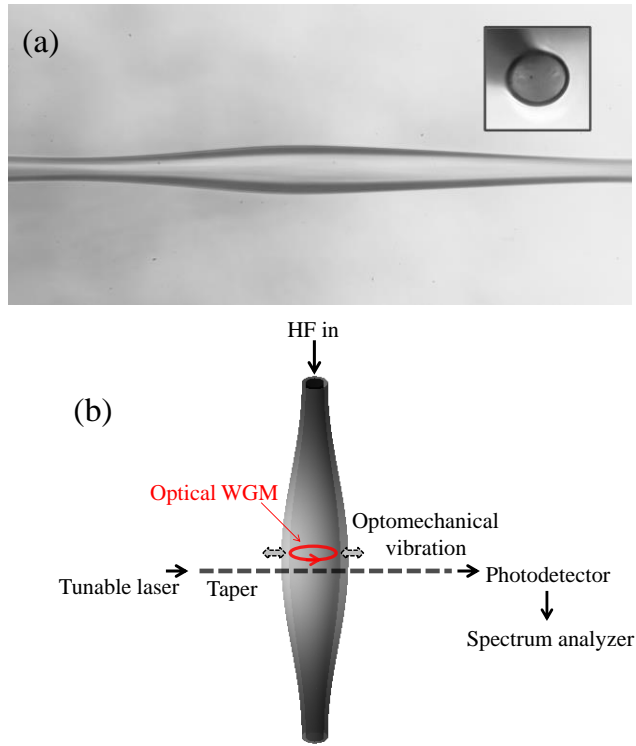

Figure 1

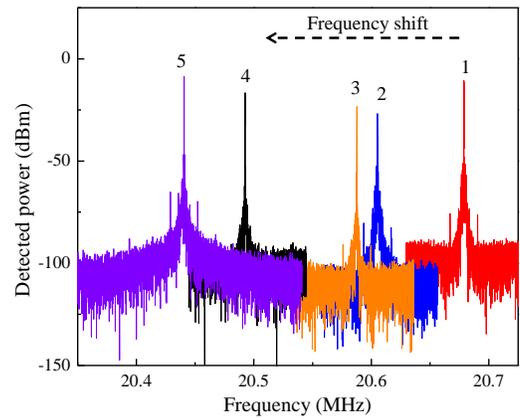

Figure 2

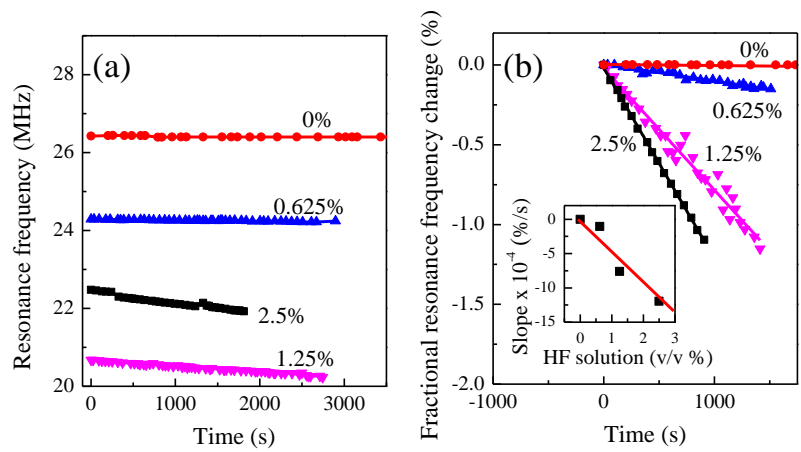

Figure 3

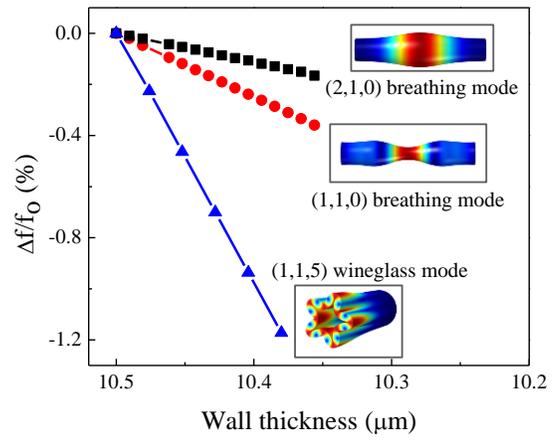

Figure 4